\documentstyle[11pt,newpasp,twoside,epsf]{article}
\def\edcomment#1{\iffalse\marginpar{\raggedright\sl#1\/}\else\relax\fi}
\reversemarginpar
\font\tenrm=cmr10
\font\ninerm=cmr9
\font\eightrm=cmr8

\begin{document}
\title{X-ray Properties of the Central kpc of AGN and Starbursts: 
        The Latest News from Chandra}
 \author{Kimberly A. Weaver}
\affil{Code 662, NASA/GSFC, Greenbelt 
MD 20771 USA\\
Department of Physics \& Astronomy, Bloomberg Center,
Johns Hopkins University, Baltimore, MD 21218, USA}

\begin{abstract}
The X-ray properties of 15 nearby ($v<3,000$ km s$^{-1}$)  
galaxies that possess AGN and/or starbursts are discussed. 
Two-thirds have nuclear extended emission on scales from $\sim0.5$
to $\sim1.5$ kpc that is either clearly associated with a nuclear outflow
or morphologically resembles an outflow.  
Galaxies that are AGN-dominated tend to have linear structures 
while starburst-dominated galaxies tend to have plume-like structures. 
Significant X-ray absorption is present in the starburst regions,
indicating that a circumnuclear starburst is sufficient to block  
an AGN at optical wavelengths.
Galaxies with starburst activity possess more X-ray point 
sources within their central kpc than non-starbursts. 
Many of these sources are more luminous than 
typical X-ray binaries.  The {\it Chandra} results are 
discussed in terms of the starburst-AGN connection, a revised 
unified model for AGN, and possible evolutionary scenarios. 
\end{abstract}

\section{Introduction}

Approximately 20\% of local galaxies possess some 
form of nuclear activity.  Such activity ranges from the 
release of gravitational energy from accretion onto a supermassive 
($\sim10^6-10^8$ M$_{\sun}$) black hole 
to the cumulative effects of short-lived episodes of 
star formation.   Active galactic nuclei (AGN)
and starburst activity often co-exist,
but because of their different observational properties, they have 
historically been studied as independent phenomena.

There are many pieces to the nuclear-activity puzzle.  
X-rays that are emitted near a supermassive black hole, whether from
the base of a jet, from the accretion disk corona or by some other
mechanism, are an excellent means of probing the matter within a few to
tens of gravitational radii ($r_{\rm g}$=GM/$c^2$) via absorption,
scattering or fluorescence in the accreting material. On larger scales, 
X-rays probe star-forming regions, starburst-driven
nuclear outflows and AGN-driven outflows.  On their journey out 
of the galaxy core, X-rays are absorbed and/or interact in other 
ways with surrounding material (torus, molecular clouds)
to produce regions such as photoionized ``warm'' absorbers and 
hot electron scattering ``mirrors''.  

There is also a new class of X-ray emitting objects: the so-called
ultra-luminous compact X-ray sources (ULXs, Makishima
et al.\ 2000) or intermediate-luminosity X-ray sources (IXOs). 
These objects have 0.1--2.4 keV luminosities higher than
the Eddington luminosity of a 1.4 solar mass accreting neutron star
($L_{\rm X} > 10^{38}$ erg s$^{-1}$).  About half of normal galaxies 
contain such sources (Colbert \& Mushotsky 1999).  Their nature
is unclear but these 
compact objects may be accreting ``middleweight'' black 
holes with masses on the order of 100 to 1,000
M$_{\sun}$ (Makishima et al.\ 2000) or 
stellar-mass black holes that are beamed toward us 
or emit anisotropically (King et al.\ 2001). 

To understand the connection between starbursts and AGN  
it is important to learn how these two phenomena
influence their common environment.  For example, 
the onset of a central starburst correlates
with a higher probability of obscuring the 
AGN broad line region (Maiolino et al.\ 1995).  This 
suggests that the presence or absence of a starburst 
is sometimes responsible for how we choose to  
classify an AGN.  Progress in this area of study requires 
the ability to disentangle the emission processes of 
circumnuclear starbursts
from AGN, observations that are now possible 
in the X-ray band with {\it Chandra}.  

\section{The data}

The X-ray data discussed here were obtained with the 
ACIS-S experiment onboard the {\it Chandra} X-ray observatory.
All but three data sets are currently in the public archive.
If published, results from the literature are reviewed.
Table 1 lists the galaxies, their NED classification 
and a secondary classification (if available) in parenthesis.

\begin{table}
\caption{{\it Chandra} Observations of Nearby Starbursts and AGN}
\vskip0.04in
\hrule \vskip0.02in  
\hrule 
\vskip0.02in
\begin{tabular}{lccrrrc}
\noalign{\vskip0.02in}
Galaxy& Type$^{(1)}$& $z^{(2)}$& $D^{(3)}$& Scale$^{(3)}$
      & ObsID$^{(4)}$& Time$^{(5)}$\\
             &  &  &(Mpc)& (pc/\arcsec)& & (ks)  \\
\noalign{\vskip0.02in}
\noalign{\hrule}
NGC 253      & Stb       & 0.0008 &  3 & 13  & 969  & 15.0   \\
NGC 1068     & S2 (Stb)  & 0.0038 & 23 & 110 & 344  & 50.0  \\
NGC 2110     & S2        & 0.0078 & 47 & 210 & 883  & 50.0  \\
MCG$-$5-23-16& S1.9      & 0.0083 & 50 & 240 & 2121 & 80.0 \\
M 82         & Stb       & 0.0007 & 4  & 19  & 361  & 34.0  \\
NGC 3079     & S2/L (Stb) & 0.0038 & 23 & 110 & 2038 & 30.0   \\
NGC 3256     & Merger(Stb) & 0.0091 & 55 & 270 & 835  & 25.0  \\
NGC 3628     & L (Stb)   & 0.0028 & 16 & 70  & 2039 & 60.0  \\
NGC 4051     & S1.5      & 0.0024 & 14 & 60  & 859  & 80.0  \\
NGC 4151     & S1.5      & 0.0033 & 20 & 90  & 348  & 30.0  \\
NGC 4258     & S1.9/L    & 0.0015 & 9  & 44  & 350  & 14.5   \\
NGC 4579     & S1.9/L    & 0.0051 & 30 &140  & 807  & 35.0   \\
NGC 4945     & S2 (Stb)  & 0.0019 & 11 & 53  & 864  & 50.0  \\
M 51         & S2        & 0.0015 & 8  & 40  & 354  & 15.0   \\
Circinus     & S2 (Stb)  & 0.0015 & 4  & 19  & 356  & 25.0  \\
\end{tabular}
\vskip0.03in
\hrule \vskip0.02in
\hrule \vskip0.04in
\ninerm{Notes.---(1) NED classification. Symbols are: S1.5, 1.9, 
2 = Seyfert classification, Stb = starburst, L = LINER.   Parenthesis
list additional classification.
(2) Redshifts are from NED. (3) Distances are calculated assuming H$_0 = 50$
km s$^{-1}$ Mpc$^{-1}$.  (4) {\it Chandra} observation ID.  
(5) Archived exposure time.}
\end{table} 

At the time of this writing, the ACIS spectral calibration was
being revised (http://asc.harvard.edu/).   
I therefore focus primarily on imaging results. 

\section{AGN and AGN-Powered X-rays} 

\begin{figure}[tbh]
{\centering \leavevmode \ninerm
\epsfxsize=0.55\textheight \hskip0.5in \epsfbox{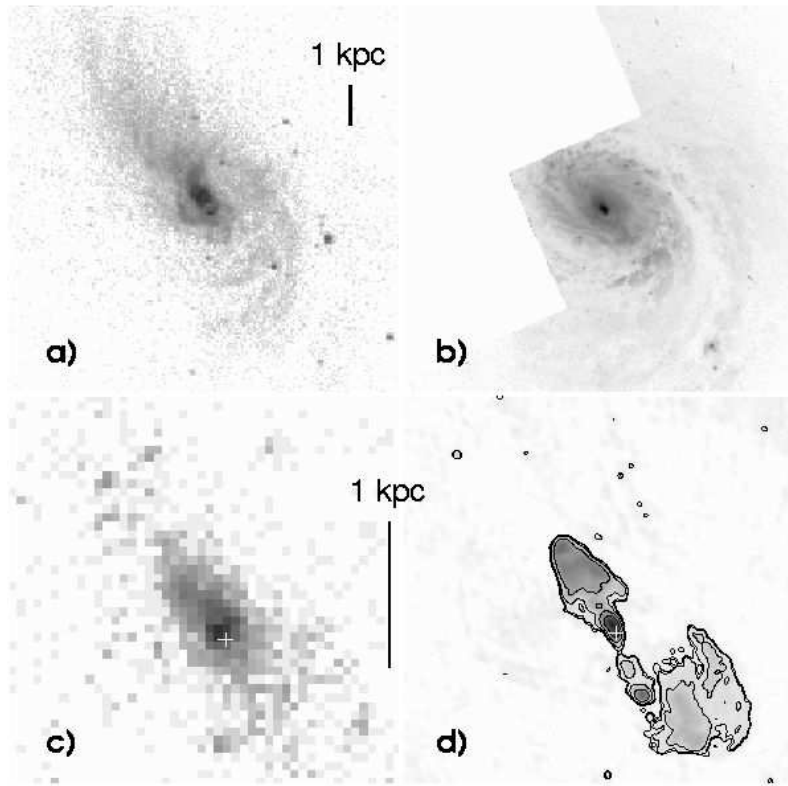} \hfill}
\caption{
\tenrm{Images of NGC 1068. {\it a)} Raw, unbinned 0.2--8.5 keV {\it Chandra}
image.  Pixels are $0.5\arcsec\times0.5\arcsec$; 
image size is $1.5\arcmin\times1.5\arcmin$ (9.9 kpc $\times$ 9.9 kpc). 
{\it b) HST} WFPC image plotted on the same scale.
{\it c)} Enlargement of the central 22$\arcsec\times22\arcsec$
(2.4 kpc $\times$ 2.4 kpc) region in hard X-rays (2.5--8.5 keV).
{\it d)} VLA 6 cm radio map
(Wilson \& Ulvestad 1983) plotted on the same scale.  
{\it Crosses} mark the position of the radio nucleus.
}}
\end{figure}

\begin{figure}[tbh]
{\centering \leavevmode
\epsfxsize=1.1\textwidth \epsfbox{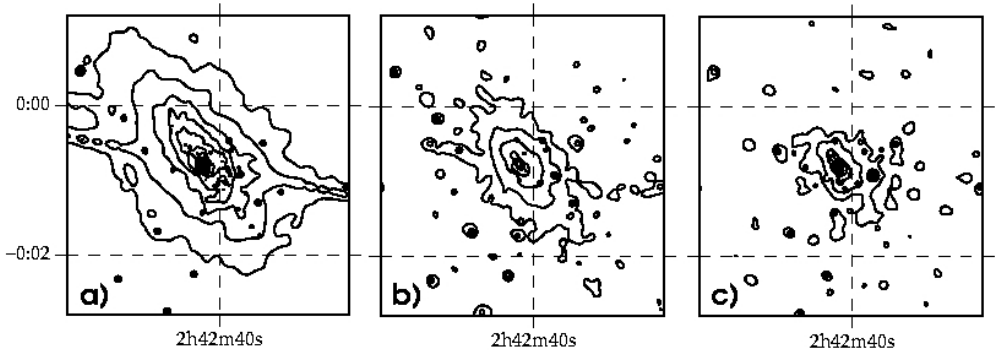} \hfil}
\vskip-0.3in
\caption{
\tenrm{Contours for
adaptively smoothed images of NGC 1068 
for energy bands {\it a)} 0.2--1.5 keV, {\it b)} 1.5--2.5 keV and 
{\it c)} 2.5--8.5 keV.  Image size is $4\arcmin\times4\arcmin$
(26.4 kpc $\times$ 26.4 kpc for H$_0 = 50$
km s$^{-1}$ Mpc$^{-1}$). 
}}
\end{figure}

\subsection{NGC 1068}
NGC 1068 is a Seyfert 2 galaxy harboring an intense starburst ring 
within its central $\sim2$ kpc.  X-ray emission covers much 
of the inner region of the galaxy (Fig.\ 1{\it a}) and shows a wispy 
structure similar to the spiral arms (Fig.\ 1{\it b}). 
Figure 1{\it c} shows an enlargement of the central region.
A bright compact
X-ray source extends $\sim1.5\arcsec$ (165 pc) in the direction of
the nuclear radio continuum emission (Young, Wilson \& Shopbell 2001).
Bright X-ray emission also extends 5$\arcsec$ (550 pc) to the NE
coinciding with the northeast radio lobe (Fig.\ 1{\it d}) and 
high-excitation ionized gas seen in [O III] $\lambda$5007 images.

The X-ray morphology in different energy bands  
offers a clue as to whether the X-ray emission
is associated with the AGN or the starburst.  
Young, et al.\ (2001) argue
that the starburst contributes little to the X-rays because
the large-scale emissions are fairly well aligned and they do not 
correlate with the $\sim2$ kpc starburst ring.
Figure 2 shows that this is clearly the case for the soft 
and medium X-ray bands, while the outer contours of the 
hard X-ray map are slightly less elliptical.  Hard X-ray emission
extends up to 2.2 kpc from the nucleus, including Fe K emission.

Using the ACIS-S data, Young, Wilson \& Shopbell (2001) rule out 
models consisting of hot plasma emission because  
such models yield implausibly low abundances. 
Instead they claim that the large-scale X-ray emission
arises from photoionization and fluorescence 
of the gas by radiation from the Seyfert nucleus. 
With a complimentary {\it Chandra} grating observation,
Ogle (2001) finds that the nuclear spectrum also contains
H-like and He-like resonance lines that are stronger than expected
for pure recombination and probably due to resonance scattering.
The off-nuclear spectrum also contains
strong resonance lines due either
to resonance scattering or thermal emission.  So a
contribution from the starburst cannot be ruled out.   

\begin{figure}[tbh]
{\centering \leavevmode
\epsfxsize=0.6\textwidth \epsfbox{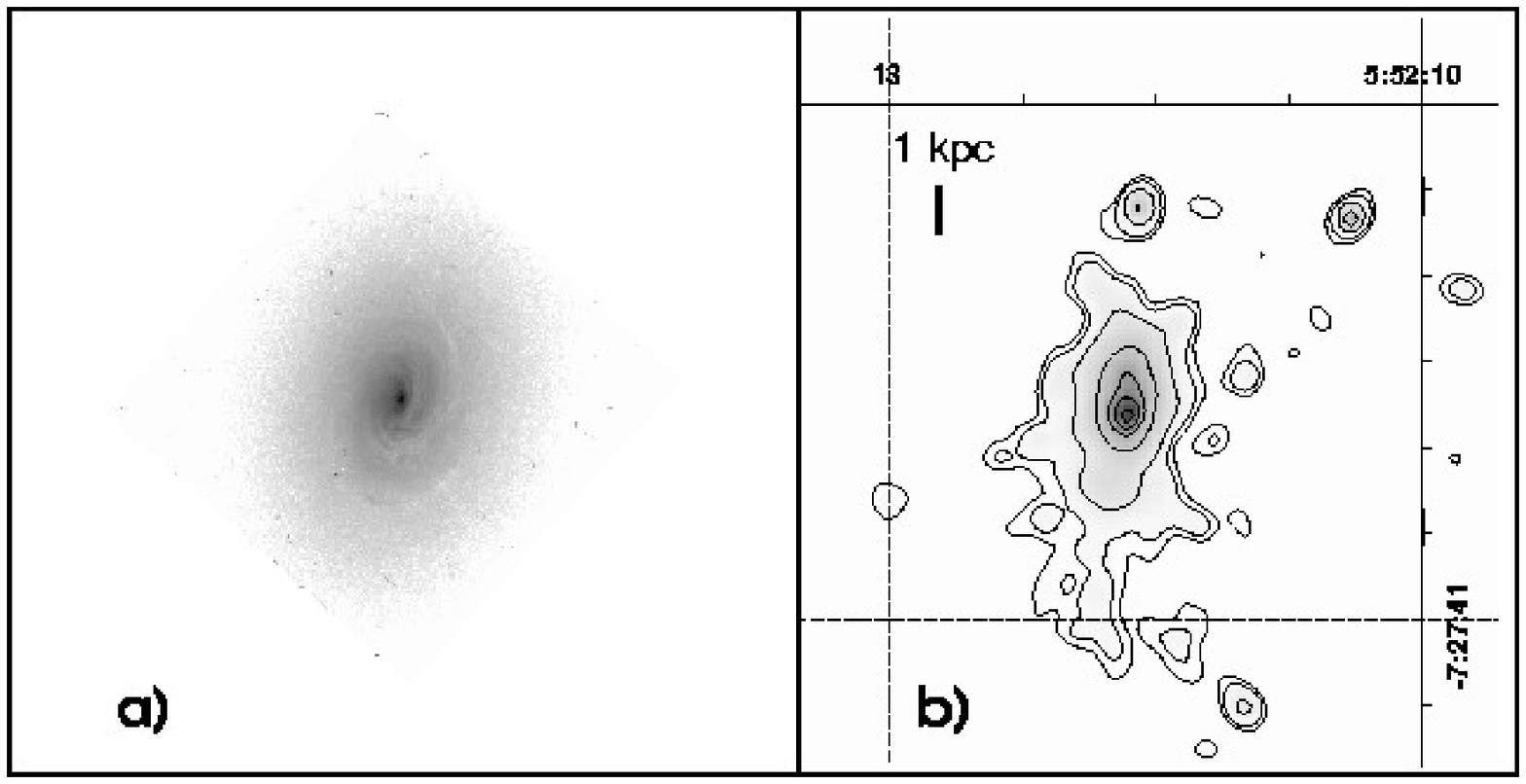} \hfil
\epsfxsize=0.35\textwidth  \epsfbox{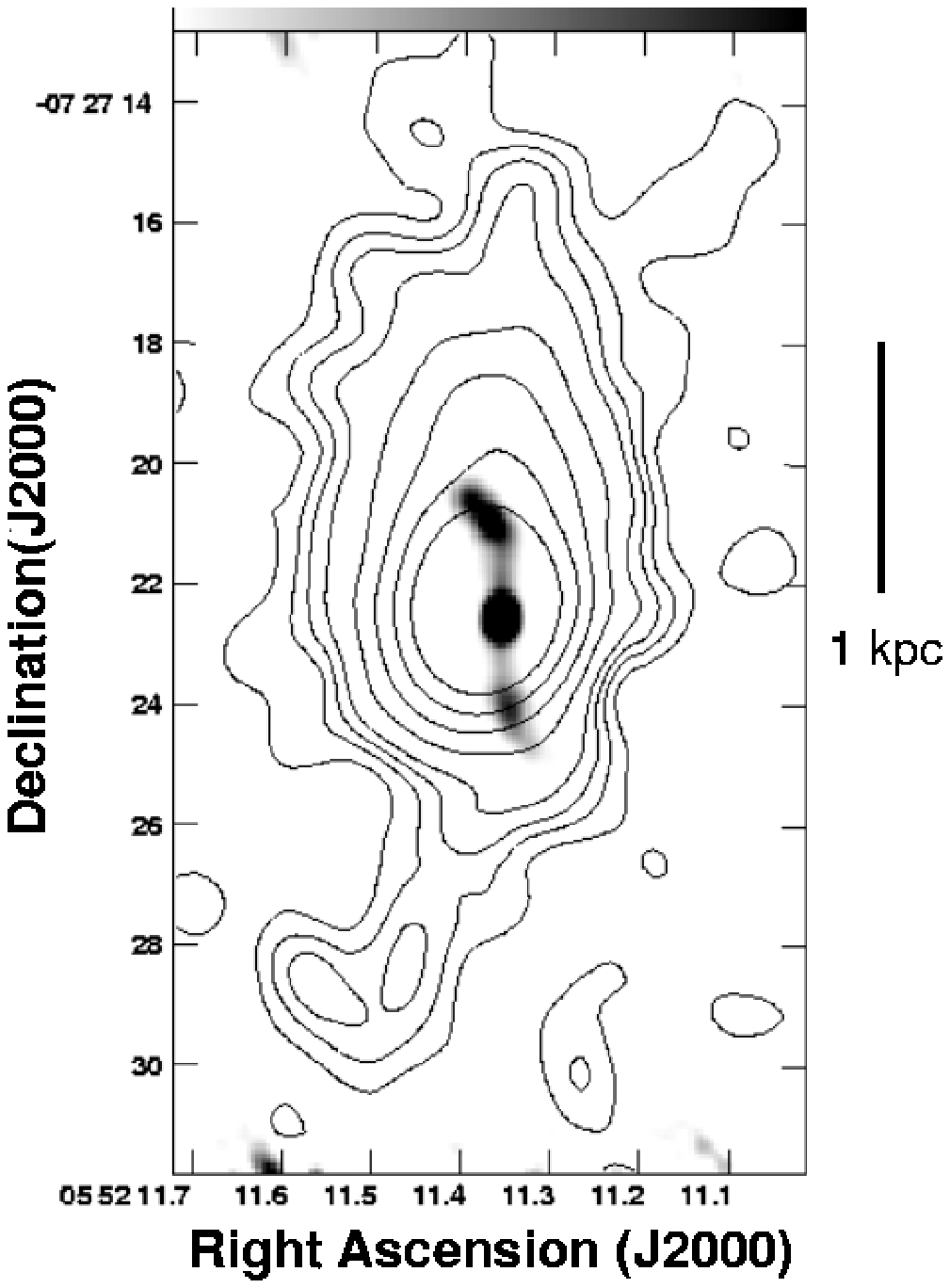} \hfil}
\caption{
\tenrm{{\it a) HST} WFPC {\it image} of NGC 2110 compared with {\it b)} the 
soft X-ray {\it Chandra} contours (0.2--1.5 keV).  The image sizes
are 60$\arcsec\times63\arcsec$ and the X-ray 
image in {\it b)} is smoothed to a resolution of 2\arcsec.
{\it c)} Enlargement of the central region showing the nuclear 
soft X-ray {\it contours} overlaid on the 6 cm radio map
(Ulvestad \& Wilson 1983).
}}
\end{figure}

\subsection{NGC 2110}
The {\it Chandra} image of NGC 2110 is shown in Figure 3.  Soft  
X-ray emission extends $\sim40\arcsec$ ($\sim$8.4 kpc)
in the direction 
of the galaxy major axis.  At the nucleus,
soft X-ray emission extends on the order of 
5\arcsec (1 kpc) along the
direction of the highest excitation optical emission line gas 
and the 6-cm radio jet (Fig. 3{\it c}).  
There are enough counts from the region surrounding the radio/X-ray
jet to obtain a crude spectrum.  The spectrum 
can be modeled with a Mekal plasma with
{\it k}T$\sim0.6$ keV and 0.05 Z$_{\sun}$ or a power law
with $\Gamma\sim4$.  The requirement for 
unphysically low abundances is similar to the
result for NGC 1068 (\S3.1).

Emission above 2 keV is spatially unresolved.
The core is absorbed with $N_{\rm H} =  
3\times10^{22}$ cm$^{-2}$ and an X-ray flux of $3\times10^{-11}$ 
erg cm$^{-2}$ s$^{-1}$.   
About \threequarters\/ of the soft X-ray emission at the nucleus can be described
as the result of partial covering of the continuum source
while about \onequarter\/ is consistent with thermal emission with a temperature 
of 0.6 keV and solar abundances.

\subsection{MCG$-$5-23-16}

The importance of the Fe K$\alpha$ line for studying the accretion phenomenon
in AGN was solidified by $ASCA$, which discovered extremely broad and asymmetric
lines (Nandra et al.\ 1997).  Observations of
MCG$-$5-23-16 with {\it ASCA} suggested that the Fe K$\alpha$ line has
a narrow core and a broad component (Weaver et al.\ 1997).
\medskip

\begin{figure}[tbh]
{\centering \leavevmode
\epsfxsize=0.38\textwidth \hskip1.5in \epsfbox{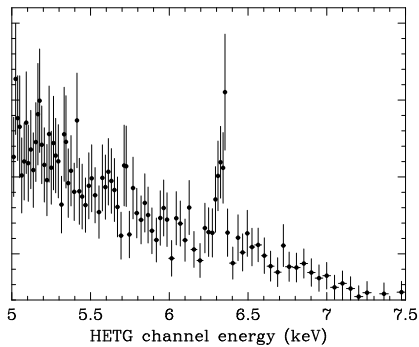} \hfil}
\vskip-0.6in
\caption{
\tenrm{HETG spectrum of the Fe K$\alpha$ line in
MCG$-$5-23-16.
}}
\end{figure}

The {\it Chandra} HETG spectrum of 
MCG$-$5-23-16 is shown in Figure 4.   The 
6.4 keV line is narrow with 
FWHM $< 3,000$ km s$^{-1}$ and an equivalent width of 
$\sim90$ eV (about \onehalf\/ the {\it ASCA} EW).  
The {\it ASCA} broad component 
is not detectable in the {\it Chandra} spectrum.
Modeling the line as emission from a geometrically thin accretion 
disk yields a lower limit to the line-emitting
region of 200 $r_{\rm g}$ (units of GM/$c^2$).  

\subsection{NGC 3079}

Figure 5 shows a false-color ACIS-S image of the soft 
X-ray emission from NGC 3079 (Strickland et al., in prep). 
The soft X-ray filaments align well with H$\alpha$ + N {\ninerm II}
emission from the wind-blown nuclear superbubble, which has 
a velocity of $\sim2,000$ km s$^{-1}$ and diameter of  
1.1 kpc (Veilleux et al. 1994).

\begin{figure}[tbh]
{\centering \leavevmode
\epsfxsize=0.45\textwidth \hskip0.3in \epsfbox{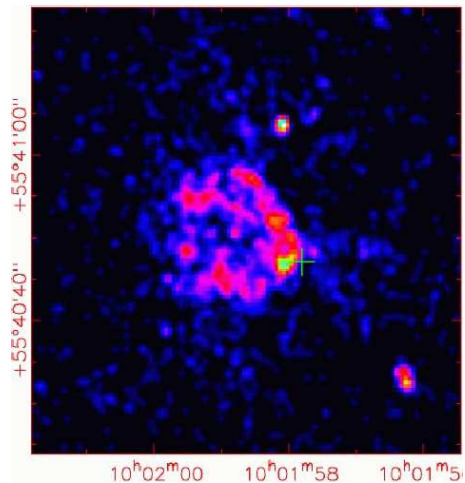} \hfill\vfill}
\caption{
\tenrm{False-color soft X-ray image of
NGC 3079.  The image is 5 kpc $\times$ 5 kpc, 
smoothed with a FWHM 1.0\arcsec\
Gaussian.  The green cross shows the position of the nucleus.
}}
\end{figure}

\subsection{NGC 4051}

Figure 6 compares the optical DSS image of NGC 4051 
with the zero order {\it Chandra} 
HETG soft and hard X-ray images.   There is no 
evidence for extended emission  
(see also Collinge et al.\ 2001).  

\begin{figure}[tbh]
{\centering \leavevmode
\epsfxsize=1.05\textwidth \hskip0.25in \epsfbox{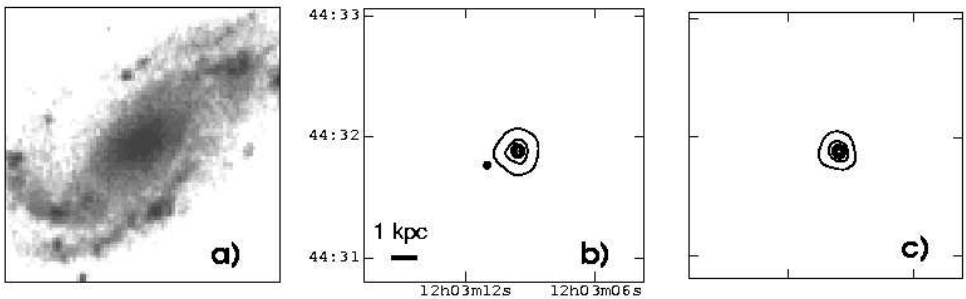} \hfil}
\vskip-0.1in
\caption{
\tenrm{Images of the central 2.4$\arcsec\times2.4\arcsec$
(8.5 kpc $\times$ 8.5 kpc) region of NGC 4051. {\it a)} Optical Digitized
Sky Survey image.  {\it b)} Contours for the smoothed {\it Chandra\/}
HETG zero-order 0.2--1.5 keV image.
{\it c)} Contours for the smoothed 2.5--8.5 keV HETG zero-order image.
}}
\end{figure}

NGC 4051 possesses an ionized absorber
with $N_{\rm H}=10^{20-21}$ cm$^{-2}$.
The {\it Chandra} spectrum shows emission lines from 
various hydrogen-like and helium-like ions of O, Ne, Mg and Si
as well as blueshifted X-ray absorption systems at 
$-2340\pm130$ km s$^{-1}$ and $-600\pm130$ km s$^{-1}$.
The high ratio of forbidden to resonance lines 
implies that the plasma is photoionized with little 
collisional ionization (Collinge et al.\ 2001).

The Fe K$\alpha$ emission line at 6.4 keV has a FWHM of less than
2,800 km s$^{-1}$ and an EW of 158 eV.  This line
is significantly narrower than an accretion disk line and, like
MCG$-$5-23-16, indicates emission from farther out
in the galaxy core.

Within 1 kpc of  
the nucleus there is a single X-ray point source, which shows 
up in the soft image to the SE of the nucleus.  It has a 0.5--8 keV
X-ray luminosity of $\sim1.4\times10^{38}$ erg s$^{-1}$, consistent 
with an X-ray binary accreting at the Eddington limit.

\subsection{NGC 4151}

\begin{figure}[tbh]
{\centering \leavevmode
\epsfxsize=0.9\textwidth \hskip0.5in \epsfbox{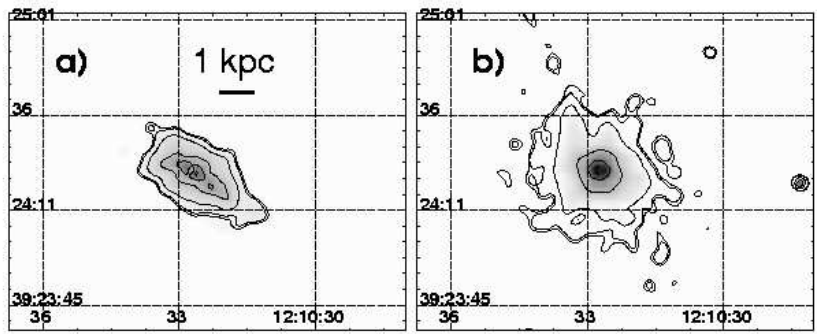} \hfil}
\caption{
\tenrm{Unbinned and adaptively smoothed greyscale
{\it Chandra\/} images of NGC 4151 with {\it contours} overlaid.
Image sizes are 1.8$\arcmin\times1.4\arcmin$ (9.7 kpc $\times$ 7.6 kpc) and are
shown for energy bands {\it a)} 0.2--1.5 keV (soft) and
{\it b)} 2.5--8.5 keV (hard).
}}
\end{figure}

About 70\% of the soft X-ray emission in NGC 4151 is
spatially resolved (Ogle et al.\ 2000) with an extent of
$\sim1.6$ kpc SW of the nucleus (Fig.\ 7{\it a}), similar to
the optical NLR emission. Whether the hard X-rays are
extended is unclear.  From ACIS-S (non-grating)
data, Yang, Wilson \& Ferruit (2001)
claim that energies $>2$ keV are spatially unresolved.
However, from grating observations,
Ogle et al.\ (2000) infer that some of the narrow Fe K$\alpha$ emission comes
from the extended NLR.

The {\it Chandra} grating spectrum is dominated by narrow emission lines. 
The X-ray NLR is composite, consisting of both photoionized
(T = $3\times10^4$ K) and collisionally ionized (T = $10^7$ K) 
components, indicating a two-phase medium
(Ogle et al.\ 2000).

\medskip
\subsection{NGC 4258}

\begin{figure}[tbh]
{\centering \leavevmode
\epsfxsize=1.1\textwidth \hskip-0.45in \epsfbox{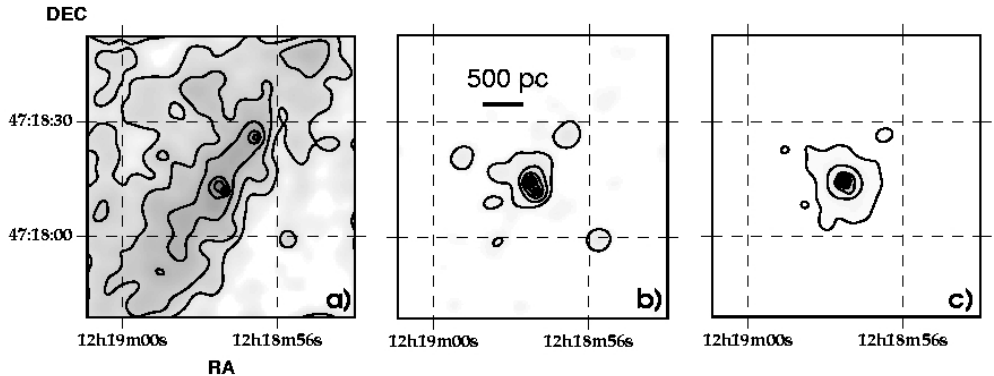} \hfil}
\vskip-0.3in
\caption{
\tenrm{Smoothed greyscale ACIS images of NGC 4258 
with {\it contours} overlaid.  Image size is 
$71\arcsec\times76\arcsec$ 
(2.5 kpc $\times$ 2.6 kpc).  Images are divided into {\it a)}
0.2--1.5 keV, {\it b)} 1.5--2.5 keV and {\it c)} 2.5--8.5 keV.
 }}
\end{figure}

{\it Chandra} images 
of NGC 4258 are shown in Figure 8.  The soft X-ray emission
is extended and trails along the ``anomalous arms''
of the galaxy.  The large-scale emission and its 
relation to the radio jets is discussed in
detail by Wilson, Yang \& Cecil (2001).   
There is little emission extended at harder energies.

There are two point sources in the core, 
with the one to the SW being brighter in soft 
X-rays (Fig.\ 8{\it a}).  The NE source 
dominates the hard X-ray image (Fig.\ 8{\it c}) and 
coincides with the radio continuum core of 
the galaxy and nuclear H$_2$O maser source (Wilson, et al.\ 2001).
This is the location of the buried AGN.  {\it SAX} 
observations measure $N_{\rm H}=9.5\times10^{22}$ cm$^{-2}$
(Fiore et al. 2001).
The SW source has properties consistent with an XRB.  

\subsection{NGC 4579}

\begin{figure}[tbh]
{\centering \leavevmode
\epsfxsize=1.06\textwidth \epsfbox{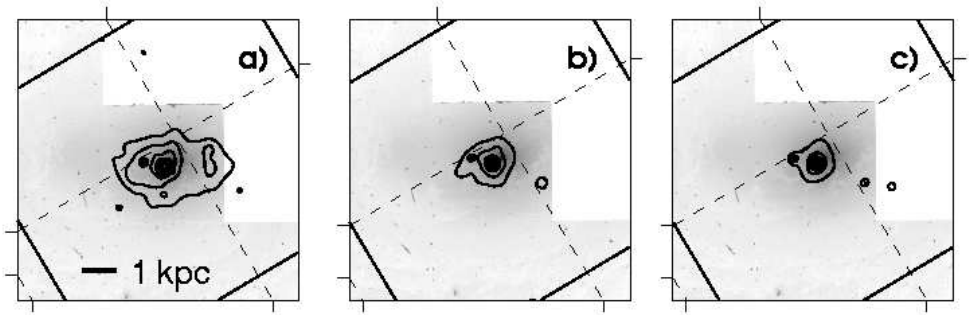} \hfil}
\vskip-0.3in
\caption{
\tenrm{ACIS-S {\it contours} overlaid on an optical {\it HST}
image of NGC 4579.   Image size is $\sim1.2\arcmin\times1.2\arcmin$ 
($\sim10$ kpc $\times$ 10 kpc).
Energy bands are {\it a)} 0.2--1.5 keV, {\it b)} 1.5--2.5 keV and {\it c)} 2.5--8.5 keV.
 }}
\end{figure}

The {\it Chandra} contours of NGC 4579 are shown 
in Figure 9, overlaid on an {\it HST image}.   The soft X-ray 
emission is extended on a scale of 
$\sim1$ kpc on either side of the nucleus. The hard X-rays are 
confined to the core, presumably from the AGN.  

There is a single off-nuclear X-ray point source about 
7$^{\prime\prime}$ from the core of the galaxy 
and possibly a few other faint point sources. 

\subsection{M 51}
{\it Chandra} images of M 51 are shown in Figure 10.
The 6 cm and 20 cm
radio contours correspond well to the extended X-ray
emission north and south of the core (Terashima \& Wilson 2001).
The northern outflow is $\sim800$ pc across and its association
with the radio emission implies shock heated gas along
the radio lobe.
The spectrum of the extended emission is well described as a
0.5--0.6 keV Mekal plasma with abundances $<0.14$ times solar.
Such temperatures indicate shock velocities of 
660--690 km s$^{-1}$. 

\begin{figure}[tbh]
{\centering \leavevmode
\epsfxsize=1.13\textwidth \hskip-0.68in \epsfbox{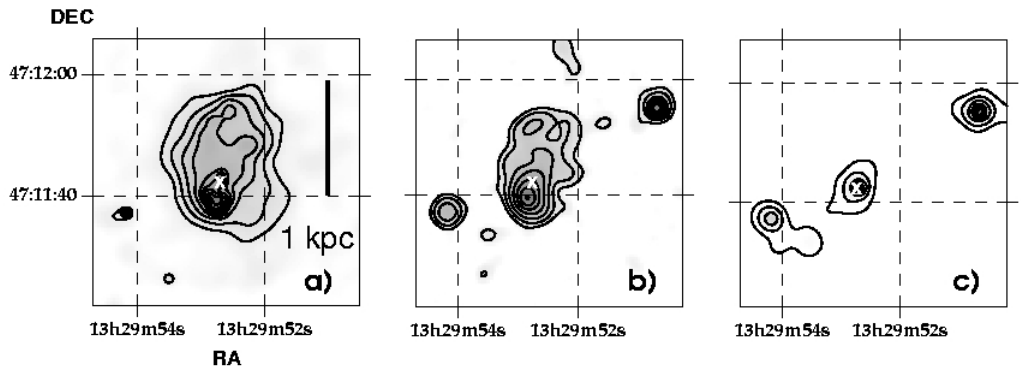} \hfil }
\vskip-0.34in
\caption{
\tenrm{Adaptively smoothed {\it Chandra} images of M51 with
{\it contours} overlaid for {\it a)} 0.2--1.5 keV, {\it b)} 1.5--2.5 keV and
{\it c)} 2.5--8.5 keV.  Image size is 1\arcmin $\times$ 1\arcmin 
(2.4 kpc $\times$ 2.4 kpc).
{\it Crosses} mark the location of the galaxy nucleus for 
comparison between images.}}
\end{figure}

The spectrum of the nucleus is dominated by reflection at
hard energies.  The best fitting model is a power law plus 
Fe K line plus Mekal plasma with {\it k}T = 0.6 keV.  
However, the plasma model prefers
low abundances, similar to NGC 1068 (\S3.1) and 
the X-ray jet in NGC 2110 (\S3.2).
For a power-law model, $N_{\rm H}=3\times10^{23}$
cm$^{-2}$.  {\it BeppoSAX} finds an additional heavily absorbed component
with intrinsic $N_{\rm H}\sim1\times10^{24}$ and an intrinsic
X-ray luminosity of $2\times10^{41}$ erg s$^{-1}$
(Fukazawa et al. 2001).

\section{Starbursts and Starburst-Powered X-rays}

\subsection{NGC 253}

Figure 11
shows the adaptively smoothed, false-color {\it Chandra} images of
the central 1.75\arcmin $\times$ 2.0\arcmin 
(1,365 pc $\times$ 1,560 pc) of NGC 253 in the
soft, medium and hard X-ray bands defined as 0.2--1.5 keV, 1.5--4.5 keV
and 4.5--8.0 keV, respectively.  
The medium and hards bands differ from those used for other galaxies 
discussed here and were chosen based on {\it ROSAT} 
and {\it ASCA} data to separate the various spectral
components (Dahlem, Weaver \& Heckman 1998).
The true-color {\it Chandra} image is shown in Figure 12{\it a}.  
The central kpc harbors a collimated, limb
brightened outflow.
The coincidence of soft X-rays with H$\alpha$ suggests 
that the X-rays in the nuclear plume are produced in
the regions of interaction between the starburst and the dense
ISM (Strickland et al.\ 2000).  

\vskip-0.1in
\begin{figure}[tbh]
{\centering \leavevmode
\epsfxsize=1.05\textwidth \epsfbox{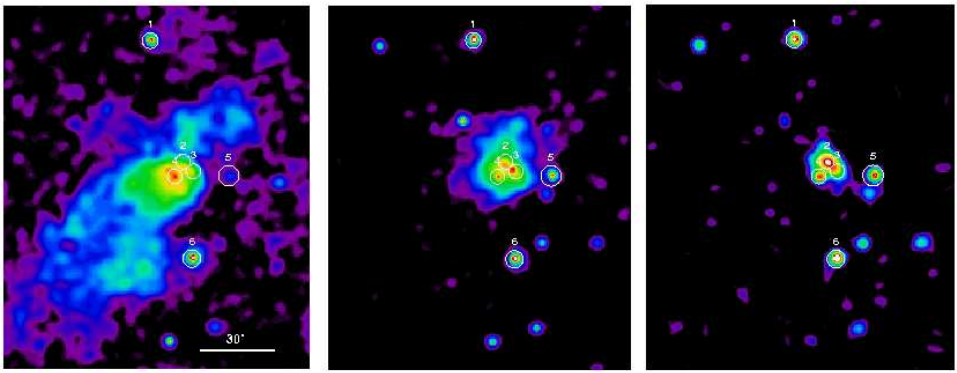} \hfil}
\vskip-0.2in
\caption{
\tenrm{Soft, medium and hard-band adaptively smoothed (false-color)
images of NGC 253.  Image size is $1.1\arcmin\times1.4\arcmin$.
Ultraluminous X-ray point sources are numbered, including the core
(\#2).}}
\end{figure}

The six brightest point sources within the 
central kpc are numbered in Figure 11.
Source 2 is coincident with the galaxy nucleus, defined by 
the brightest compact radio source (Turner and Ho 1985).  
Regions of soft and hard X-ray emission have 
different morphologies.   There is extended hard 
X-ray emission at the galaxy core associated with a 
$\sim100$ pc ridge of molecular clouds 
defined by IR and radio observations.
The numbered point sources all 
have luminosities at least 10 times larger than expected for
the Eddington Limit of a $\sim 1$ M$_{\sun}$ accreting
compact object. 

\begin{figure}[tbh]
{\centering \leavevmode
\epsfxsize=0.85\textwidth \hskip0.5in \epsfbox{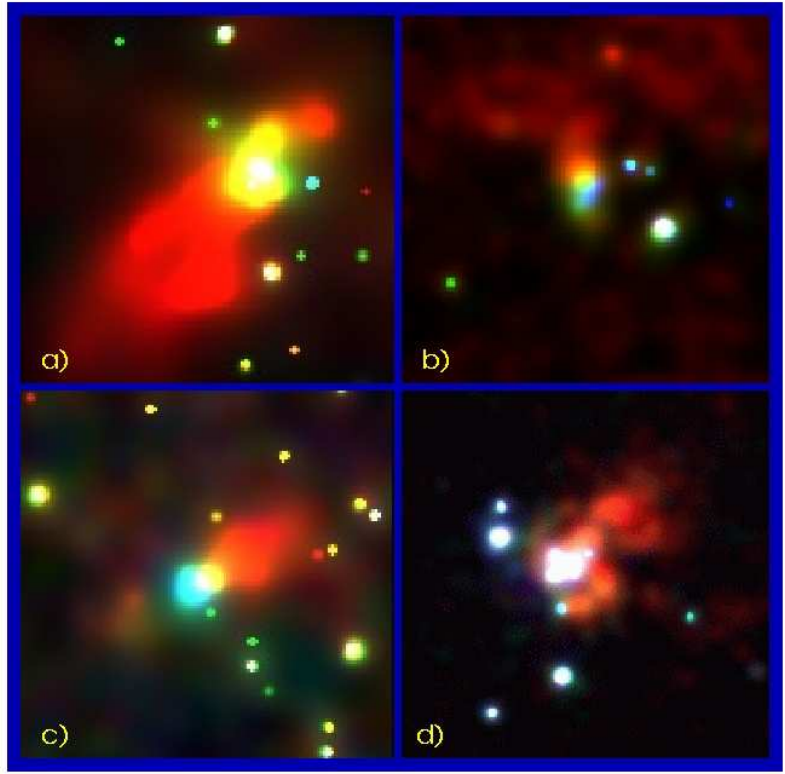} \hfill}
\caption{
\tenrm{Adaptively smoothed ``true-color'' images of {\it a)} the central
0.9 kpc $\times$ 0.9 kpc region of NGC 253, {\it b)} the central
4 kpc $\times$ 4 kpc region
of NGC 3628, {\it c)} the central 2.5 kpc $\times$ 2.5 kpc region of NGC 4945
and {\it d)} the central 1.6 kpc $\times$ 1.6 kpc region of Circinus.
For NGC 3628,
red is 0.3--1.0 keV, green 1--2 keV, blue 2--8 keV.
For the other galaxies, red is 0.2--1.5 keV, green 1.5--2.5 keV,
blue 2.5--8.5 keV.
}}
\end{figure}

The core 
has an unabsorbed $0.2-9$ keV flux of $6.7\times10^{-13}$ 
erg cm$^{-2}$ s$^{-1}$, $N_{\rm H}=3\times10^{22}$ cm$^{-2}$ and
an unabsorbed luminosity of $2\times10^{39}$ erg s$^{-1}$.
It possesses a hard X-ray spectrum ($\Gamma \sim 1.3$) 
with copious line emission. 
A thermal model yields a temperature of 12 keV, which is
consistent with what might be expected for hot gas in the starburst region
(Chevalier \& Clegg 1985); however, the emission lines
are much stronger relative to the continuum than expected for 
a thermal plasma, suggesting that the continuum source 
is obscured from view.
If the central source is buried and completely absorbed
(requiring $N_{\rm H} \ge10^{24}$), then
the power-law continuum likely represents scattered X-rays from the central
source reflected from the molecular clouds.  In this case,
the intrinsic nuclear luminosity is probably 
close to $1\times10^{41}$ erg s$^{-1}$.

\subsection{M 82}

\begin{figure}[tbh]
{\centering \leavevmode
\epsfxsize=0.5\textheight \hskip0.7in \epsfbox{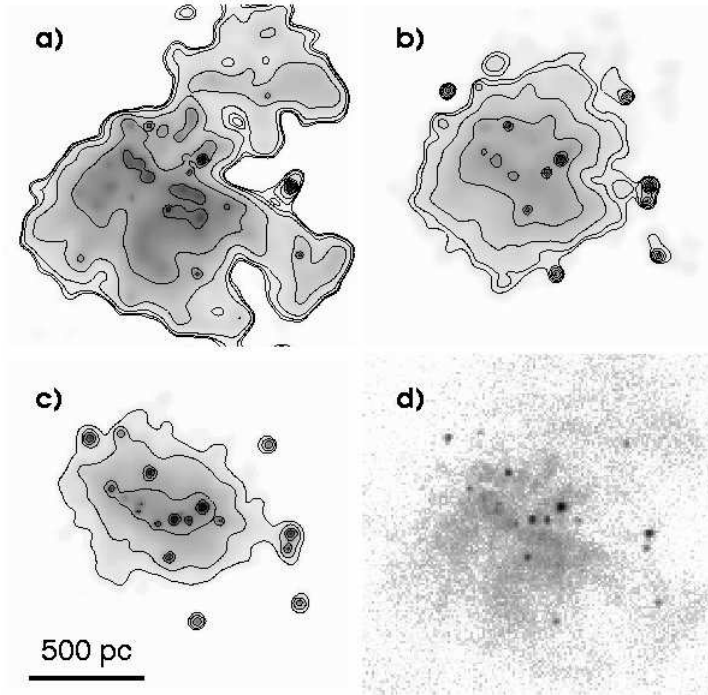} \hfil}
\caption{
\tenrm{{\it a--c)} Adaptively smoothed {\it Chandra} ACIS-S greyscale
images of the central 1.25$\arcmin\times1.25\arcmin$ 
(1.4 kpc $\times$ 1.4 kpc)
region of M82 with {\it contours} overlaid.  Energy bands are {\it a)}
0.2--1.5 keV , {\it b)} 1.5--2.5 keV and {\it c)} 2.5--8.5 keV.
{\it d)} Unsmoothed (0.5$\arcsec\times0.5\arcsec$
pixels) full-band image (0.2--8.5 keV).}
}
\end{figure}

{\it Chandra}
images of M 82 are shown in Figure 13.
Within the central kpc, at least 11 distinct point sources are seen,
some with significant variability (Matsumoto et al.\ 2001).
However, unlike NGC 253, there is no compact X-ray source located at the
center of the galaxy (Kaaret et al.\ 2001).  The most luminous 
point source lies 9\arcsec ($\sim170$ pc) from the kinematic center
of the galaxy (Matsumoto et al.\ 2001). Its variability
places an upper limit on its size of 0.08 pc (Kaaret et al.\ 2001)
and its luminosity ranges from about $10^{40}$ to $10^{41}$ erg s$^{-1}$,
which corresponds to the Eddington luminosity for a 500 to
900 M$_{\sun}$ object.

The central kpc region is filled with diffuse
hot gas, extending $\sim800$ pc in the 2--8 keV band
(Griffiths et al.\ 2000 and Fig.\ 13{\it c}).  
The diffuse emission makes up about
30\% of the total hard X-ray emission in the galaxy core 
and is at least partly thermal in origin, depending on
assumptions about the physical state of the gas.
M 82 has an unusually high temperature plasma core
(T $\sim4\times10^7$ K).

\subsection{NGC 3628}

The true-color, smoothed {\it Chandra} image 
of NGC 3628 is shown in Figure 12{\it b}.
The blue dot at the center
of the image represents the galaxy nucleus.
The soft X-ray emission to the
north is located on the top of the disk while there is strong
absorption to the south.

The white circle is a well-known, variable
ULX (or IXO) with an X-ray luminosity of
$\sim1\times10^{40}$ erg s$^{-1}$.  It is located $900\pm25$ pc from
the nucleus in projection (Strickland et al.\ 2001).

\section{Starburst/AGN Composite Galaxies}

\subsection{NGC 3256}
NGC 3256 is a merger-induced starburst system (Moran, Lehnert \&
Helfand 1999) with an X-ray luminous starburst having a luminosity 
of $1.6\times10^{42}$ erg s$^{-1}$.  {\it ASCA} data indicate that 
the soft X-rays are primarily thermal in origin, 
with {\it k}T = 0.3 and 0.8 keV, while the hard X-rays 
are best described as a power law.  It has been suggested that 
NGC 3256 harbors two nuclei, one being a buried AGN 

\begin{figure}[tbh]
{\centering \leavevmode
\epsfxsize=0.4\textheight \hskip0.65in \epsfbox{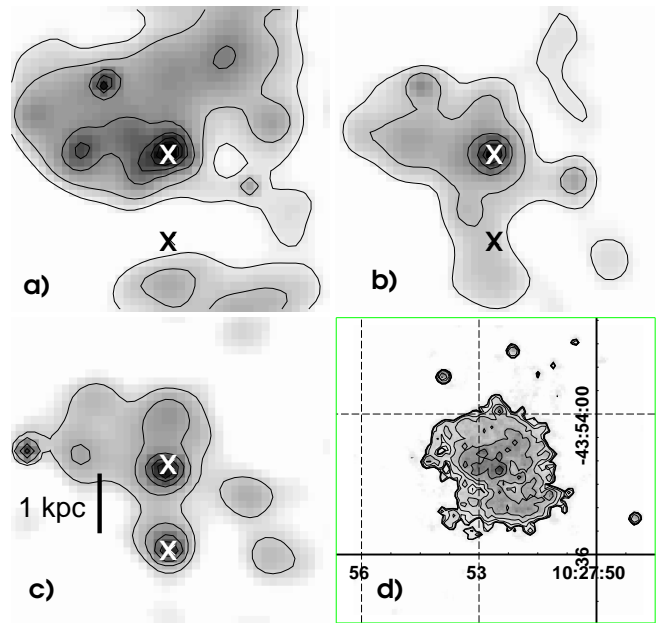} \hfil }
\caption{
\tenrm{Grey-scale ACIS-S images of NGC 3256 with {\it contours}
overlaid.  The first three panels are 20$\arcsec\times20\arcsec$
(5.4 kpc $\times$ 5.4 kpc) and represent the following energy bands:
{\it a)} 0.2--1.5 keV {\it b)}, 1.5--2.5 keV and {\it c)} and 2.5--8.5 keV.
The {\it lowest contours} are drawn approximately at the level
of the local background.
{\it d)} Adaptively smoothed full-band (0.2--8.5 keV) image
with {\it contours} overlaid.}
}
\end{figure}

The {\it Chandra} image (Fig.\ 14) shows diffuse emission
extended throughout the disk of the galaxy.
At soft energies there is a single nuclear source, but at 
hard energies, two nuclei are clearly visible.  The crosses
mark the position of the optical nucleus (which has an 
IR and radio counterpart) and an IR and radio 
point source that is offset from 
the optical nucleus by about 1 kpc 
(Kotilainen et al.\ 1996). The two IR and radio
nuclei are of comparable strength.  The Chandra image 
shows that this is true in hard X-rays as well, 
confirming the possibility of a buried AGN with $N_{\rm H}
\sim10^{23}$ cm$^{-2}$.

\begin{figure}[tbh]
{\centering \leavevmode
\epsfxsize=0.5\textheight \hskip0.75in \epsfbox{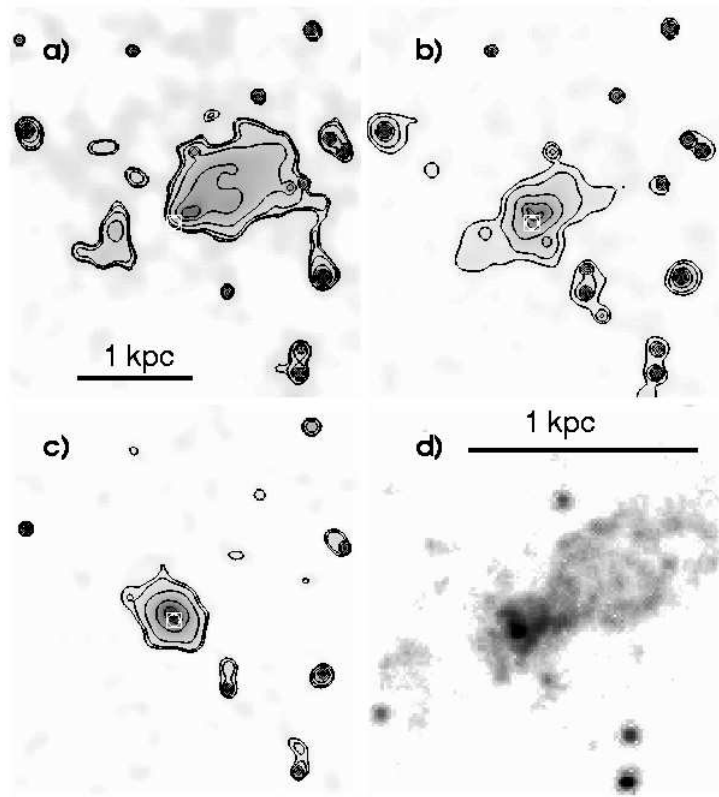} \hfil }
\caption{
\tenrm{Adaptively smoothed {\it Chandra} ACIS images of
NGC 4945 with {\it contours} overlaid.  The first three
panels are 1.0\arcmin $\times$ 1.2\arcmin (3.2 kpc $\times$ 3.8 kpc); the fourth
is 30\arcsec$\times$36\arcsec.  Images are shown for energy
bands {\it a)} 0.2--1.5 keV, {\it b)} 1.5--2.5 keV, {\it c)} 2.5--8.5 keV and
{\it d)} 0.2--8.5 keV (total).  The {\it white circle-in-square point}
marks the position of the hard X-ray continuum peak for cross
comparison of the images.}
}
\end{figure}

\subsection{NGC 4945}
NGC 4945 shows a soft X-ray ``plume'' that lies perpendicular
to the galaxy disk and extends $\sim1$ kpc from the nucleus
(Fig.\ 15 and Fig.\ 12{\it c}).  Like NGC 253, the soft X-rays 
suffer heavy absorption from the galaxy disk.  The hard X-ray 
contours are slightly elongated in the direction of the galaxy disk
which suggests some contribution from the starburst. 
The spectrum of the plume is
dominated by soft X-ray lines from a combination
of photoionized and collisionally ionized plasma.
The nuclear spectrum is dominated by 
the Fe K line which presumably comes from a small ($<100$ pc) 
region (G. M. Madejski, private communication).

\begin{figure}[tbh]
{\centering \leavevmode
\epsfxsize=0.58\textheight \hskip0.5in \epsfbox{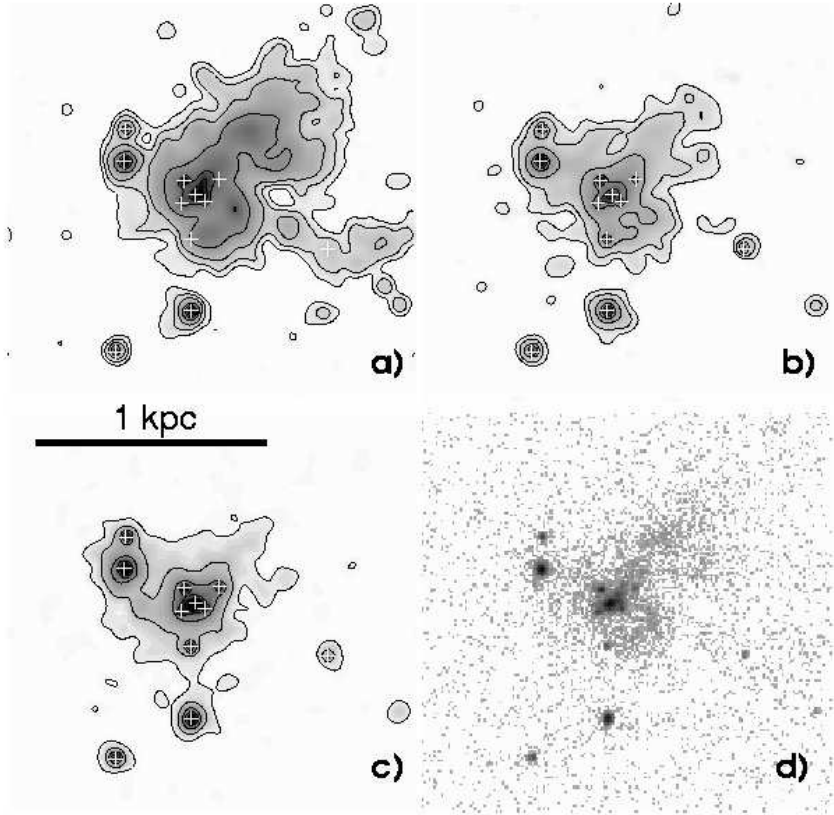} \hfil }
\caption{
\tenrm{{\it a--c)} Adaptively smoothed ACIS-S images with {\it contours} 
for the central 1.4\arcmin $\times$ 1.4\arcmin\
(1.6 kpc $\times$ 1.6 kpc) region of Circinus.  Energy bands are  
{\it a)} 0.2--1.5 keV, {\it b)} 
1.5--2.5 keV and {\it c)} 2.5--8.5 keV.  {\it Crosses} mark 
the positions of the point sources.
{\it d)} Unbinned and unsmoothed full-band image (0.2--8.5 keV).}  
}
\end{figure}

\subsection{Circinus}

The soft X-ray emission from Circinus has an
asymmetrical extension to the northwest in 
the direction of the optical ionization cone.
The soft X-ray ``plume'' (Fig.\ 16 and Fig.\ 12{\it d})
coincides with the large-scale [O III] emission
(Smith \& Wilson 2001).  The
X-rays may be associated with a hot radiatively driven wind,
shocks, or highly excited photoionized gas.  Closer
to the core on smaller scales ($\sim50$ pc), soft X-rays
are consistent with scattered nuclear radiation by ionized gas
(Sambruna et al.\ 2001).   The line emission is a 
a combination of highly ionized material together with a
low-ionization reflector.  

Eleven point sources are embedded
in the diffuse X-ray emission.  Bauer et al.\ (2001) report that few 
have optical counterparts within 1\arcsec of their
X-ray positions down to limiting magnitudes of $m_{\rm V}$=23--25.
The brightest is identified with the galaxy nucleus.
At the distance of Circinus, their intrinsic
0.5--10 keV luminosities range from $\sim2\times 10^{37}$
to $\sim4\times10^{39}$ erg s$^{-1}$ with seven having 
L$_X$ in excess of $1.3\times10^{38}$ erg s$^{-1}$.  
One-fourth of the point sources
vary and their spectral properties suggest that most 
are either black hole binaries or  
ultraluminous supernova remnants.

The ACIS spectrum of the
nucleus exhibits emission lines at both soft and hard X-rays, 
including a prominent (EW $\sim2.5$) Fe K$\alpha$ line at 6.4 keV.
Fe K$\alpha$ emission 
extends up to 200 pc from the nucleus (Smith \& Wilson 2001).

\section{Results Summary}

Table 2 summarizes the spectral/spatial results from
{\it Chandra} observations.  For cases where results are not
available in the literature (or current calibration issues
preclude deriving them), results from other X-ray
experiments such as $ASCA$ and {\it BeppoSAX} are listed in italics.

\begin{table}
\caption{X-ray Spatial and Spectral Results}
\vskip0.05in
\hrule \vskip0.02in
\hrule \vskip0.02in
\begin{tabular}{lcccccc}
Galaxy & Nuclear &  \# pt. src$^{(1)}$ & Nuclear$^{(2)}$
          & $N_{\rm H}$ & Ref.$^{(3)}$ \\
           &  Extent &  within  &
             Spectrum & ($10^{22}$) & \\
     &   &  1 kpc  &   &   &  \\
\noalign{\hrule}
\noalign{\vskip0.02in}
NGC 253    & $\sim1$ kpc & 7    & M,P,c         & 3               &  1 \\
NGC 1068   & 550 pc      & 4    & P,M,R,F,p,c   &0.1,{\it 1000}  &  2,3,{\it 4} \\
NGC 2110   & $\sim1$ kpc & 0    & P,M,F         & 3               &  \\
MCG$-$5-23-16&$<1$ kpc   & --   & P,F           & 5               &   \\
M 82       &  1 kpc      & 11   & M,F,c         & $\sim1$         &  5,6,7 \\
NGC 3079   &  1.1 kpc    & 1    &{\it M,P} &{\it ~1}       & {\it 8} \\
NGC 3256   & --          & 2    &{\it M,P} &{\it 0.8}, 20  & {\it 9} \\
NGC 3628   & $\sim1$ kpc & 3    &{\it M,P} &{\it 0.9}      & 10, {\it 8}\\
NGC 4051   &$<400$ pc    & 1    & P,F,p         &0.1              &  11 \\
NGC 4151   &$\sim1.5$ kpc& 1    & P,F,p,c       & 3               &  12,13 \\
NGC 4258   & --          & 2    & P,F          & $\sim1$,{\it 10} &  14,{\it 15} \\
NGC 4579   & --          & 2     & --           & --               &   \\
NGC 4945   & $\sim1$ kpc & 4     & P,F,p,c     & {\it 400}      & {\it 16} \\
M 51       & 800 pc      & 1    & R,M,F,c &3, {\it 600}    &  17, {\it 18}  \\
Circinus   &60,600 pc    & 10   & R,M,F,p,c    & 0.4,{\it 400}  & 19,20,21,{\it 4}\\
\end{tabular}
\hrule \vskip0.02in
\hrule
\vskip0.1in
\eightrm{{\it Notes}.---(1) The number of X-ray point sources within 1 kpc of the nucleus,
excluding the nucleus itself.

(2) Dominant continuum spectral components.  P=power law, M=Mekal plasma,
R=Compton reflection.  Whether emission lines appear to be produced 
primarily from p=photoionization
or c=collisional ionization.  F indicates that an Fe K line is present.}

\ninerm{(3) References: 1) Strickland, et al.\ 2000,
2) Young, et al.\ 2001, 3) Ogle 2001, 4) Bianchi, et al.\ 2001, 5)
Griffiths, et al.\ 2000, 6) Matsumoto, et al.\ 2001, 7) Kaaret, et al.\ 2001,
8) Dahlem, et al.\ 1998, 9) Moran, et al.\ 1999,
10) Strickland, et al.\ 2001, 11) Collinge, et al.\ 2001, 12)
Ogle, et al.\ 2000, 13) Yang, et al.\ 2001, 14) Wilson, et al.\
2001, 15) Fiore, et al.\ 2001.,
16) Madejski, et al.\ 2000, 17) Terashima \& Wilson 2001, 18) Fukazawa, et al.\ 2001,
19) Sambruna, et al.\ 2001, 20) Bauer, et al.\ 2001, 21) Smith \& Wilson 2001.}
\end{table}

\subsection{Extended Emission}

All of the galaxies mentioned in this review, with 
the exception of NGC 4051 and MCG$-$5-23-16, possess extended 
soft X-ray emission.  At least three galaxies, namely 
NGC 1068, M 82 and Circinus, have significant extended hard  
X-ray emission.
Ten of the 15 galaxies possess a 
distinct nuclear component of extended soft X-rays
that resembles an outflow.
The morphology of this nuclear extended emission 
falls into two categories, ``plume-like'' structures  
and linear ``jet-like'' structures.

Galaxies with X-ray plumes all possess starbursts.  
These are NGC 253, NGC 4945, NGC 3079, NGC 3628 and Circinus.
M 82 is also plume-like although the nuclear emission is not 
as easily distinguished from the larger scale X-ray emission.
The plumes range in size
from $\sim600$ pc to $\sim1$ kpc and their spectra 
tend to be dominated by thermal emission. 

Galaxies with X-ray jets are NGC 1068, NGC 2110,
and NGC 4151.  The X-ray features align with optical and 
radio counterparts and their sizes range from $\sim500$ pc
to $\sim1.5$ kpc.  The X-ray jets tend to have fairly 
featureless spectra; their spectra are not well described 
as thermal emission.  The nuclei of galaxies with X-ray 
jets all harbor AGN and their nuclear spectra tend 
to be dominated by photoionization.

M 51 is unique in that its X-ray outflow resembles a 
cross between a jet and a plume.  The X-rays have 
a radio counterpart, but the X-ray spectrum is thermal. 
This is perhaps the best case for shock heating resulting 
from an AGN-driven outflow. 

The presence of an X-ray plume appears to depend on 
the presence of a starburst.  Two of the most 
spectacular plumes occur in NGC 253, a starburst,
and NGC 4945, a starburst with an AGN.  It would 
appear that the collimating mechanisms are the same
in these two nuclei.  While it is fairly clear 
that the plume in NGC 253 is a superwind, in NGC 4945
it is not clear whether the outflow represents a 
superwind or an ionization cone.  By analogy with 
NGC 253, it seems likely that 
the starburst is driving the 
outflow in NGC 4945 and that, in this case, the outflow 
is also the ionization cone.

\subsection{The X-ray Spectra}

Complex absorption structures are common in AGN and starbursts.
The starburst region can have large X-ray column densities
of up to a few times $10^{22}$ cm$^{-2}$.  This is similar to the 
column densities seen in X-ray bright Seyfert 2s like NGC 2110
and MCG$-$5-2-16.   But there are also multiple regions or layers of 
absorption in many of these galaxies.  Buried AGN are often hidden 
behind absorbing columns a factor of 10 to 100 times higher than those 
measured with {\it Chandra} (Table 2).
Examples are M 51, NGC 4945, NGC 1068 and Circinus.

Ten galaxies show prominent 
Fe K$\alpha$ lines.  
MCG$-$5-23-16 and NGC 4051 have lines that   
are dominated by a narrow and unresolved  
core that cannot arise from the inner regions 
of an accretion disk.
Fe K emission also is extended on NLR scales in 
at least two cases (Circinus and NGC 1068).
Clearly, non-disk contributions to the Fe K$\alpha$ 
emission in Seyfert galaxies are common. 

\subsection{Other X-ray point Sources Within the Central kpc}

The galaxies with the largest number of 
X-ray point sources within their central kpc 
(excluding the actual nucleus) are 
galaxies with starbursts like NGC 253, M 82 and Circinus. 
Pure AGN have the fewest (0 to 2), while the other galaxies 
tend to have 1 to 4.  It is important to examine more 
nearby galaxies to rule out a selection effect since 
the galaxies with the most point sources are 
closest to us and hence individual sources are easier 
to detect. However,
in NGC 253 and Circinus, many of the point sources are 
very bright and ultraluminous and similar sources could have been 
detected in at least some of the other galaxies.  This suggests that   
near-nuclear ultraluminous sources are more 
common in galaxies with circumnuclear  
starbursts.

\section{The Starburst-AGN Connection}

I believe that the results presented here support the idea of an 
intimate connection between starburst and AGN activity.
{\it Chandra} images and spectra offer powerful evidence 
for the interplay between stellar and
non-stellar activity in the centers of nearby galaxies. 
The starburst region can have large X-ray column densities
of up to a few times $10^{22}$ cm$^{-2}$.  Such large 
absorption can obscure the
central AGN at optical wavelengths, making 
a Seyfert 1 galaxy in to a Seyfert 2
without requiring a pc-scale molecular torus.  A similar effect has 
been inferred for a large $ASCA$ sample 
of AGN with starbursts (Levenson, Weaver \& Heckman 2001).

\begin{figure}[tbh]
{\centering \leavevmode
\epsfxsize=0.6\textheight \hskip0.7in \epsfbox{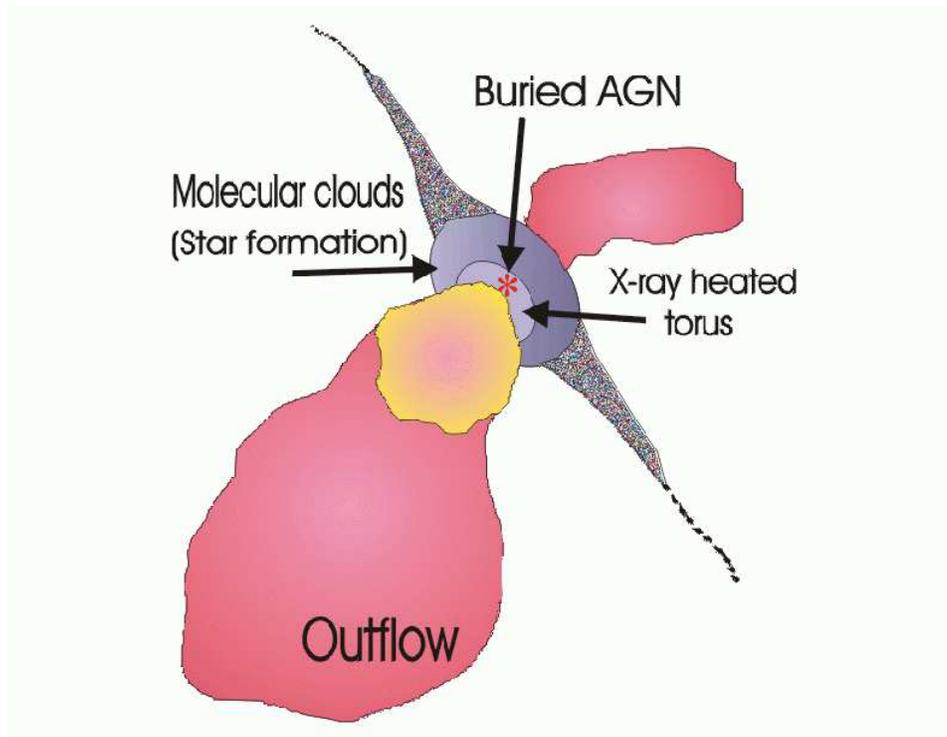} \hfill\vfill}
\caption{
\tenrm{
Cartoon of the central kpc of a starburst-AGN
composite (``energetic'') galaxy.
}}
\end{figure}

A physical model for the center of an energetic galaxy is 
proposed in Figure 17, which shows a cartoon of 
an AGN buried within a circumnuclear starburst.  If active, the AGN  
will photoionize the inner edge of the ring of molecular clouds 
associated with the starburst, creating a large-scale 
X-ray heated torus.  The emergent nuclear spectrum would 
be complex, showing features of reflection,
ionized absorption, photoionization,  
collisional excitation and fluorescence, 
similar to many of the {\it Chandra} spectra. 
This problem is similar to that of X-ray heated winds in AGN
(Krolik \& Kriss 1995) but adding a secondary source 
of ionization and/or heating from the starburst. 

Figure 18 shows a revised version of
the standard unified model for AGN. 
On the pc scale is the Compton-thick 
molecular torus, needed to provide the heavy obscuration
of the central source in objects like NGC 1068, NGC 4945,
M 51.  At distances of $\sim$50--100 pc are the
starburst region clouds with column
densities of $10^{22-23}$ cm$^{-2}$, enough to obscure X-ray
bright Seyfert 2s like NGC 2110, MCG$-$5-23-16
(a starburst ``curtain'').
Depending on the alignment of the Seyfert
nucleus with the starburst geometry,
the ionization cone could be blocked  
or even further collimated by the starburst.   

\begin{figure}[tbh]
{\centering \leavevmode
\epsfxsize=0.9\textwidth \hskip0.28in \epsfbox{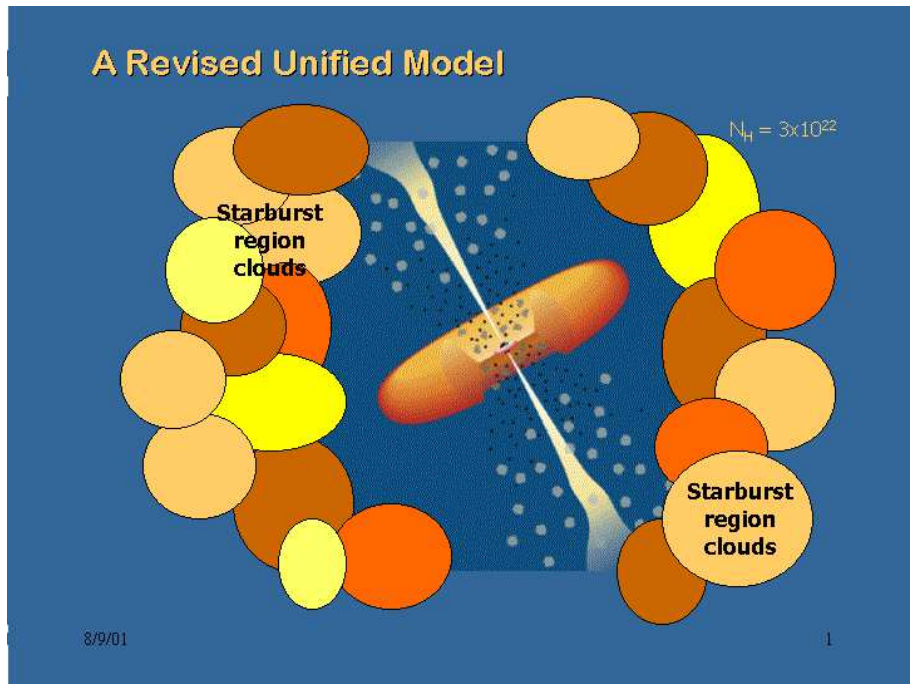} \hfill}
\caption{
\tenrm{Cartoon of a revised unified model, which includes a central
AGN surrounded by the standard pc-scale Compton-thick torus plus an
obscuring region of $\sim50$ pc diameter of dense molecular
clouds ($N_{\rm H}=10^{22-23}$ cm$^{-2}$) associated
with a circumnuclear starburst.}
}
\end{figure}

It has been proposed that circumnuclear starburst
activity and nuclear activity could both be triggered by gas 
accreting toward the nucleus.  This gas could also be 
responsible for obscuring the AGN in its early stages of
formation (Sanders et al.\ 1988).  Indeed,   
Oliva et al.\ (1999) find that old and
powerful starbursts are common in obscured AGN.
It has also been suggested that a starburst could evolve into an  
AGN by building up a massive star cluster at the 
center of a galaxy which then triggers an AGN by feeding a 
central black hole as a result of mass loss (Norman \& Scoville 1988).
Galaxies may cycle through phases 
where the starburst outshines the AGN and vice versa.

Finally, bright off-nuclear X-ray point sources (some 
being ULXs with possible masses of 10--1,000 M$_{\sun}$)
appear to be common in circumnuclear starburst regions.
It is possible that such ULX black holes (if they are 
indeed intermediate mass black holes) 
may be the precursors to AGN activity.
If born in dense star clusters near the centers of galaxies, ULX BHs
could sink to the core via dynamical friction, eventually growing
into a supermassive black hole (Tremaine, Ostriker \& Spitzer 1975;
Quinlan \& Shapiro 1990).

\acknowledgments
Thanks to David Strickland, Patrick Ogle and Greg Madejski
for communicating results prior to publication.
This research has made use of the NASA/IPAC Extragalactic Database (NED) 
which is operated by the Jet Propulsion Laboratory, California Institute 
of Technology, under contract with the National Aeronautics and
Space Administration. 
The Digitized Sky Survey was produced at the Space Telescope Science Institute under
U.S. Government grant NAG W-2166. The images of these surveys are based on
photographic data
obtained using the Oschin Schmidt Telescope on Palomar Mountain and the UK
Schmidt Telescope.

\end{document}